\documentclass[prl,twocolumn,superscriptaddress,showpacs]{revtex4-1}

\usepackage{amsmath}
\usepackage{amssymb}
\usepackage{graphicx}
\usepackage[noxcolor]{pstricks}
\usepackage{hyperref}

\def\ba{\begin{eqnarray}}
\def\ea{\end{eqnarray}}
\def\nl{\nonumber \\}

\newcommand{\ZZ}{\mathbb{Z}}

\begin{document}

\title{Evaluation of ranks of real space and particle entanglement spectra for large systems}


\author{Iv\'{a}n D. Rodr\'{i}guez}
\affiliation{Department of Mathematical Physics, National University of Ireland, Maynooth, Ireland}

\author{Steven H. Simon}
\affiliation{The Rudolf peierls Centre for Theoretical Physics, 1 Keble Road, Oxford, OX1 3NP, united Kingdom}

\author{J.K. Slingerland}
\affiliation{Department of Mathematical Physics, National University of Ireland, Maynooth, Ireland}
\affiliation{Dublin Institute for Advanced Studies, School of Theoretical Physics, 10 Burlington Rd, Dublin, Ireland}


\begin{abstract}
We devise a way to calculate the dimensions of symmetry sectors appearing in the Particle Entanglement Spectrum (PES)  and Real Space Entanglement Spectrum (RSES) of multi-particle systems from their real space wave functions. We first note that these ranks in the entanglement spectra equal the dimensions of spaces of wave functions with a number of particles fixed. This also yields equality of the multiplicities in the PES and the RSES. 
Our technique allows numerical calculations for much larger systems than were previously feasible. For somewhat smaller systems, we can find approximate entanglement energies as well as multiplicities. We illustrate the method with results on the RSES and PES multiplicities for integer quantum Hall states, Laughlin and Jain composite fermion states and for the Moore-Read state at filling $\nu=\frac{5}{2}$, for system sizes up to $70$ particles.   
\end{abstract}

\date{\today}


\maketitle

In recent years, entanglement has been embraced as an important tool in the study of many particle systems.
After splitting the system into subsystems $A$ and $B$ the entanglement between these can be quantified using the entanglement entropy $S_A=-\rm{tr}(\rho_{A}\log(\rho_{A}))$, where $\rho_{A}$ is the density matrix of subsystem $A$. One splitting is obtained by dividing the space in which the system resides into disjoint parts. For topological phases of matter, one finds that $S_A$ then increases in proportion to the size of the boundary between these, up to a constant topologically invariant correction, the \emph{topological entanglement entropy} $\gamma$~\cite{Kitaev06_PRL,Levin06}. In an attempt to calculate $\gamma$ for fractional quantum Hall (QH) states, the authors of Refs.~\onlinecite{Haque07,Zozulya07} introduced a splitting of Hall systems in angular momentum space, the \emph{orbital cut}. Since the orbitals of the Landau problem localize the particle near rings or lines whose location relates to the angular momentum, one may view the orbital cut as a cut in real space. However, as spatial regions of high probability density can overlap strongly for different orbitals, the orbital cut is actually very `fuzzy' in real space. A different cut which simply divides the particles in two groups, irrespective of their positions, was also considered in Refs.~\onlinecite{Iblisdir07,Haque07,Zozulya07}. We will see that, perhaps counterintuitively, this cut relates more directly to a sharp spatial cut.
In Ref.~\onlinecite{Li08}, Li and Haldane introduced the \emph{entanglement spectrum} as a generalization of the entanglement entropy. They consider all eigenvalues of $\rho_A$, and the corresponding states in the Schmidt decomposition.  Li and Haldane conjectured a correspondence between the states in the orbital entanglement spectrum (OES) of the ground state of a Hall system and the edge excitations of one of the subsystems. This implies that one may make study of the edge excitations of many gapped systems even if only the ground state of the system can be accessed directly. A similar conjecture~\cite{Sterdyniak11_PRL} states that the particle entanglement spectrum (PES) should reproduce the spectrum of the smaller subsystem at the same total flux as the full system. 
These correspondences between energy and entanglement spectra are necessarily qualitative, since relevant data like edge potentials are not represented in the bulk ground state. Also, in model systems, some classes of excitations may not be reflected in the ground state - one may think of fully spin-polarized systems which have excitations involving spin flips. Nevertheless, 
generic arguments in favor of the OES-edge correspondence were presented in Refs.~\onlinecite{Qi11} and~\onlinecite{Swingle11} in cases where the edge has a conformal field theory (CFT) description. In Ref.~\onlinecite{Chandran11}, it was also shown that, for a large class of Hall states, the PES contains the same information about the edge as the OES in the thermodynamic limit. Moreover, the correspondence holds for systems beyond QH states, including topological insulators and superconductors~\cite{Turner09,Fidkowski10}.

Here, we calculate the angular momentum multiplicities of the \emph{real space entanglement spectrum} (RSES), i.e. the ES which results from a sharp cut in real space (as in the definition of the topological entanglement entropy). These multiplicities are the ranks of the angular momentum blocks of the reduced density matrix. For QH states, the RSES and RSES ranks have so far been calculated only for the non-interacting states at some integer fillings~\cite{Rodriguez09,Rodriguez10}. The method of calculation that we develop works just as well for the PES ranks
and we will soon see that the RSES and PES ranks are in fact equal. This means in particular, using the results of Ref.~\onlinecite{Chandran11}, that, in many cases, the RSES ranks contain the same information about the thermodynamic limit of the edge as the OES ranks, though finite size effects are different. In general, we expect the RSES to give a better description of the edge than the OES. For example, for the integer Hall states,
the RSES ranks reproduces the edge multiplicities faithfully, while the OES is trivial.



\begin{figure}[hbt]
\begin{center}
\vspace*{-7mm}
\includegraphics[width= 8.5cm, height=5cm]{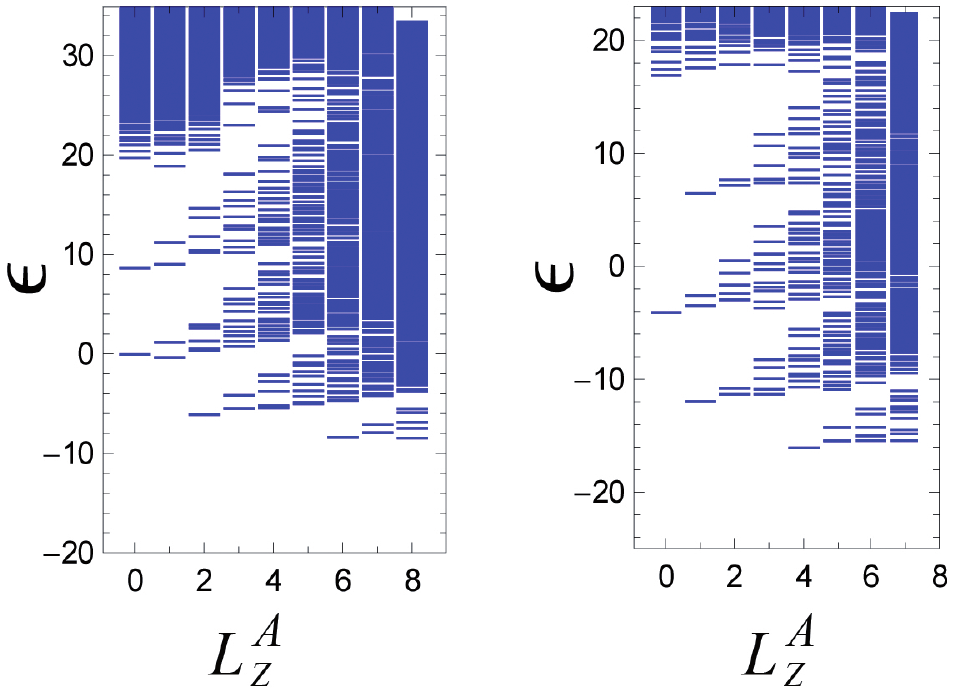}
\caption{Plot of the spectra of the matrices $M^{\dagger}M$ used in the calculation of the RSES ranks for a $\nu=\frac{2}{5}$ Jain state on a sphere at $N=20$ and $N_A=10$ (left) and at $N=22$ and $N_A=11$ (right). Negative values of $\epsilon$ occur because we used an unnormalized Jain state (normalization shifts the vertical axis by a constant). The counting $1,2,5,10,\ldots$ of the branches is consistent with a pair of noninteracting Luttinger liquids. Differences in total counting for even vs.~odd $N_A$ arise analogously to those for $\nu=2$ (cf. Eq.~(\ref{eq:nu2count})).}
\label{fig:Jain25}
\end{center}
\end{figure}

Consider a system of $N$ particles in its ground state $\psi$.
We start with a general statement of the conjectured correspondence between the entanglement spectrum and energy spectrum of such a system.
Assume that the system has a symmetry $\hat{J}=\sum_{i=1}^{N}{\hat{j}_i}$ and that $\psi$ is an eigenstate of $\hat{J}$. 
If the splitting of the system preserves $J$, then the reduced density matrices of the subsystems will be block diagonal, with the blocks labeled by the eigenvalues of $J_A$ (or $J_B=J-J_A$). It is natural to expect that the number of (low lying) states in the entanglement spectrum obtained from the ground state $\psi$, with a given eigenvalue of $J_A$, will be the same as the number of (low energy) states of subsystem $A$ which have that same eigenvalue of $J_A$. One may further hope that the entanglement energies and the real energies of the low-lying states become proportional in the thermodynamic limit and that the Schmidt states approach the low lying states in the energy spectrum. Here, we focus mainly on the ranks of the blocks of $\rho_A$, but we do present preliminary results on entanglement energies (see Fig.~\ref{fig:Jain25}). 


Now take $\psi(z_1,..,z_N)$ to be a QH wave function (with $z_i$ the complex coordinates of the particles) and take $J$ to be the angular momentum component $L_z$. The subsystems $A$ and $B$ contain $k$ and $N-k$ particles. The matrix elements of the $L_{z}^{A}$ blocks of $\rho_{A}$ are then
%
\begin{equation}
\begin{array}{l}
\rho_{A}^{(L^A_z)}(z_{1}\ldots z_{k},z'_{1}\ldots z'_{k}) = 
\int  \prod_{j=k+1}^{N} d z_{j}\, \\
~~~\int d\phi\, e^{-2\pi i L^A_z \phi}
\psi( z_{1}e^{2\pi i\phi}\ldots z_{k}e^{2\pi i\phi}, z_{k+1} \ldots z_{N})\\
~~~\int d\phi'\, e^{2\pi i L^A_z \phi'}
\psi^{*}( z'_{1}e^{2\pi i\phi'}\ldots z'_{k}e^{2\pi i\phi'}, z_{k+1} \ldots z_{N}). 
\end{array}
\label{eq:rho_A3}
\end{equation}
%
Here the role of the $\phi$ and $\phi'$ integrations is to enforce that the angular momentum in subsystem $A$ equals $L_z^A$, therefore selecting a single block of $\rho_{A}$. 
Defining $Z_A=(z_1,..,z_k)$, $Z_B=(z_{k+1},..,z_N)$  and
\ba
&& \Phi(L^A_z, Z_A,Z_B) = \nl && \int d\phi\, e^{-2\pi i L^A_z\phi} \psi( z_{1}e^{2\pi i\phi},..,z_{k}e^{2\pi i\phi},z_{k+1},..,z_{N}), 
\label{effective_psi}
\ea
we can rewrite (\ref{eq:rho_A3}) as follows:
\ba
\!\!\! \!\!\! \rho_{A}^{(L^A_z)}(Z_A;Z'_A) = \!\!\!
\int \!\! d Z_B  \Phi(L^A_z,Z_A,Z_B)  \Phi^*(L^A_z,Z'_A,Z_B). 
\label{eq:rho_final}
\ea
If the positions in $Z_A$ can take any value and the integrals over $Z_B$ are over all space, then the above expressions are for the reduced density matrix with the particle cut. However, we can restrict the $Z_A$ positions to lie in a subset of space (e.g.~a disk) and take the integrals over $Z_B$ to be over the complement of that subset. The same expressions then represent density matrix blocks for the real space cut. We can even allow the regions containing the $Z_A$ and $Z_B$ positions to overlap, giving cuts which interpolate between particle cut and real space cut.



The right hand side of (\ref{eq:rho_final}) is the 
Gram matrix for a set ${\bf P}$ of wave functions for the particles in subsystem $B$: 
\ba 
{\bf P} = \{ \tilde{\Phi}_{L^A_z,Z_A}(Z_B) = \Phi(L^A_z,Z_A,Z_B)\}. 
\label{set_qh}
\ea
Therefore, the rank of $\rho_A$ is equal to the number of linearly independent wave functions in ${\bf P}$. The wave functions in ${\bf P}$ are precisely the $L_{z}^{B}=L_{z}^{\psi}-L_{z}^{A}$ components of the wave functions for a system of $N_{B}$ electrons in a background of $N_A$ fixed electrons at positions $Z_A$, or if you will, for $N_{B}$ electrons with $N_A$ `electron holes' at positions $Z_A$. While the positions $Z_A$ are continuous, the number of linearly independent wave functions of this kind is finite as long as the Hilbert space for $N_{B}$ particles contains only a finite dimensional subspace at the given angular momentum. This is true for wave functions built from a finite number of orbitals, e.g. from orbitals in a finite number of Landau levels.
Much stronger bounds on the ranks hold when $\psi$ satisfies vanishing properties when some number of electron coordinates coincide (e.g.~if $\psi$ is the bosonic Moore-Read (MR) Pfaffian~\cite{Moore91}, it vanishes whenever $3$ particles coincide). Since all wave functions in ${\bf P}$ satisfy the same vanishing properties, the ranks are bounded by the numbers of independent states with the given vanishing properties at the given magnetic flux.  

We now argue that the ranks of the blocks of $\rho_A$ are the same for the particle cut and real space cut. 
In both cases the ranks equal the numbers of independent wave functions in ${\bf P}$. The functions in ${\bf P}$ actually remain the same, only the domains on which they are defined are different; in one case the positions $Z_A$ and $Z_B$ can be anywhere in space and in the other case, they are restricted to two disjoint regions. However, up to geometrical factors, these functions are polynomials. If two polynomials are equal on any open domain, then by analytic continuation, they must be equal everywhere. Therefore the number of independent states is the same for particle cut and real space cut. We should stress that despite the equality of ranks, the actual eigenvalues of $\rho_A$ will be very different for the two cuts and there is physical information contained in these. For example for the RSES we expect lower entanglement energies for states associated with the edge. In fact, we could choose the cut so that subsystems $A$ and $B$ share multiple circle boundaries and in that case we would expect that each edge will give rise to a branch of low entanglement energy states in the RSES. 
 
Given trial wave functions for quasiholes, we can attempt a direct proof of the correspondence between the number of independent wave functions for quasiholes and the entanglement spectrum rank by expanding the wave functions $\tilde{\Phi}_{L^A_z,Z_A}$ in an appropriate polynomial basis. If the same count is obtained for the $\tilde{\Phi}_{L^A_z,Z_A}$ as for quasihole wave functions at the same flux, then the correspondence is proved. Note that the quasihole wave functions which should be compared with the $\tilde{\Phi}_{L^A_z,Z_A}$ will have $\frac{e}{e^{*}}N_{A}$ quasiholes, where $e^{*}$ is the charge of a single quasihole. 
We will not pursue this here. Instead, we develop a numerical technique for the calculation of the RSES and PES ranks, which can be applied for any $\psi$. 


Our basic strategy is as follows. 
Even though the set of index configurations $Z_{A},Z_{A'}$ is infinite, the rank of $\rho_{A}^{(L_{z}^{A})}$ is finite. Therefore we can find it by evaluating the rank of a suitably large submatrix which we obtain by choosing a particular set of values of $Z_A$ and $Z_{A'}$. We will use a square $d\times d$ submatrix with the same set of $Z_A$ and $Z_{A'}$, which we call $Z_{A}^{i}$ with $i\in{1,\ldots,d}$. The rank of this matrix is the same as the number of linearly independent wave functions in the set 
\ba 
{\bf \tilde{P}} = \{ \tilde{\Phi}_i(Z_B) = \Phi(L^A_z,Z_{A}^{i},Z_B) \quad i=1,..,d \}. 
\label{set_qh_1}
\ea
This number is equal to the rank of the matrix
\ba
M_{i j} =  \tilde{\Phi}_i ( Z_B^j)  \qquad i=1,\ldots,d,~j=1,\ldots,d'\ge d
\label{eq:li_states}
\ea
where $Z_B^j = (z_{k+1},..z_N)_j \ (j=1,..,d'$) is a set of $d'\ge d$ different $(N-k)$-tuples of coordinates in the B subsystem. The rank of this matrix can be obtained (e.g.~by singular value decomposition) and equals the number of nonzero eigenvalues of $M^{\dagger}M$. In numerical calculations, all eigenvalues of $M^{\dagger}M$ will be nonzero, but a clear jump between large and small (nearly zero) eigenvalues is observed 
and hence the true rank can be read off. 

The scheme just described will not work without a judicious choice of
the index configurations $Z_A^i$ and $Z_B^j$. Choosing these at random will lead to very
small and greatly varying values of the matrix elements $M_{ij}$. This
induces numerical error which makes it difficult to
identify a clear cut in the spectrum of eigenvalues of $M^{\dagger}M$.
A better set of index configurations can be obtained by Monte Carlo
sampling $\psi$, but harvesting only configurations whose $Z_{A}$
satisfies a suitable constraint, which helps to select index
configurations which have good overlaps with the desired angular
momentum sector. The constraint is introduced by means of the relationship between the coordinates and the angular momentum in the Landau levels. For instance on the disk the single particle state with angular momentum $l$ has most of the probability density concentrated near a ring of radius $\sqrt{l}$. Therefore, we can assign to each Metropolis configuration $ \tilde{Z}_A  = ( z_1,..,z_k )$ a $k$-tuple of angular momenta $l_A=(l_{A}^{1},..,l_{A}^k)$ by taking $l_{A}^i$ equal to the integer nearest to $|z_i|^2$. We then harvest only those $Z_A$ which satisfy $\sum^k_{i=1} l_{A}^i  = L^A_z$ as index configurations.

When two configurations $Z_A$ and $Z'_A$ obtained in this way share the same $l_A$, we can discard one of them without loss of rank, because the corresponding rows of the matrix $M$ are almost exactly proportional. 
Moreover, to obtain the full rank of $\rho_A^{(L_z^A)}$ it is not necessary to generate index configurations corresponding to each admissible $k$-tuple $l_A$. This is because the set of states labeled by the $Z_{A}$ is overcomplete and states with different values of $l_A$ have nonzero overlaps. 
To obtain the full rank $r$ of one of the $\rho_A^{(L_z^A)}$, we typically need to take a number of indices $d$ which is only a few times larger than $r$. If we want to focus on the low lying part of the RSES, a further simplification takes place. Now the electron coordinates in the $Z_{A}^{i}$ must also lie inside the spatial region associated with subsystem $A$. E.g.~the $k$ particles in subsystem $A$ may be located in a disk of radius $r_A$, so that 
$|z_i| \le r_A$ for $i \in\{1,..,k\}$. In this case the associated angular momenta $l_{A}^{i}$ must also have their probability density concentrated on rings located within this disk and the number of $l_{A}$ which satisfy the constraint that $\sum^k_{i=1} l_{A}^i  = L^A_z$ is much smaller than if $z_1,\ldots,z_k$ could be located anywhere. This selection of indices allows for a very efficient evaluation of the RSES ranks in the angular momentum sectors associated with the edge, but it does make it more difficult to find the multiplicities for other values of $L_z^A$ associated with bulk excitations. To obtain those it is better to select indices $Z_{A}^{i}$ with electron coordinates which vary through all space. 
 
Using this method, we can manage systems with up to $N=100$ and $N_A=50$ particles on a standard laptop. Note that for systems where the correspondences between the ES and the bulk and boundary excitations hold, this gives us information on the excitations for systems of up to $50$ particles, which is considerably larger than the system sizes accessible by exact diagonalization (typically no more than $20$ particles).
With more effort and resources, much larger systems should be accessible.
It should also be clear the method can be used on the plane, the cylinder and the torus.

%


We now present a summary of results of sample PES and RSES rank calculations. These were done for integer filling fractions on the cylinder and for the Moore-Read Pfaffian~\cite{Moore91}, Laughlin and Jain states both on the cylinder and in the spherical geometry~\cite{Haldane83}. We consider the case $N_A=N/2$ and for the RSES we define the subsystem $A$ as half of the sphere or cylinder.

For the Laughlin states at $\nu=1/2$ and $\nu=1/3$ we calculated the RSES ranks for systems of $70$ particles. The multiplicities of the low-energy edge excitations of the Laughlin state, in the thermodynamic limit, are predicted by the edge conformal field theory (CFT), which is a chiral Luttinger liquid. They are
\ba
\begin{array}{|r|cccccccccc|}
\hline
         \Delta L_z  & 0 & 1 & 2 & 3 & 4 & 5 & 6 & 7  &  8  & n \\ 
\mbox{edge counting} & 1 & 1 & 2 & 3 & 5 & 7 & 11 & 15 & 22 & p(n) 
\\ \hline 
\end{array}
\label{count_Laugh}
\ea
Here $\Delta L_z$ is the relative angular momentum between the quasihole excitations and the Laughlin groundstate and $p(n)$ is the number of partitions of 
$n$.
We have computed the RSES ranks up to $L^A_z = 20$ and find that for both $\nu=1/2$ and $\nu=1/3$, they match the edge counting above.  
%
%
\begin{figure}[thb]
\begin{center}
\includegraphics[width= 7.5cm, height=5cm]{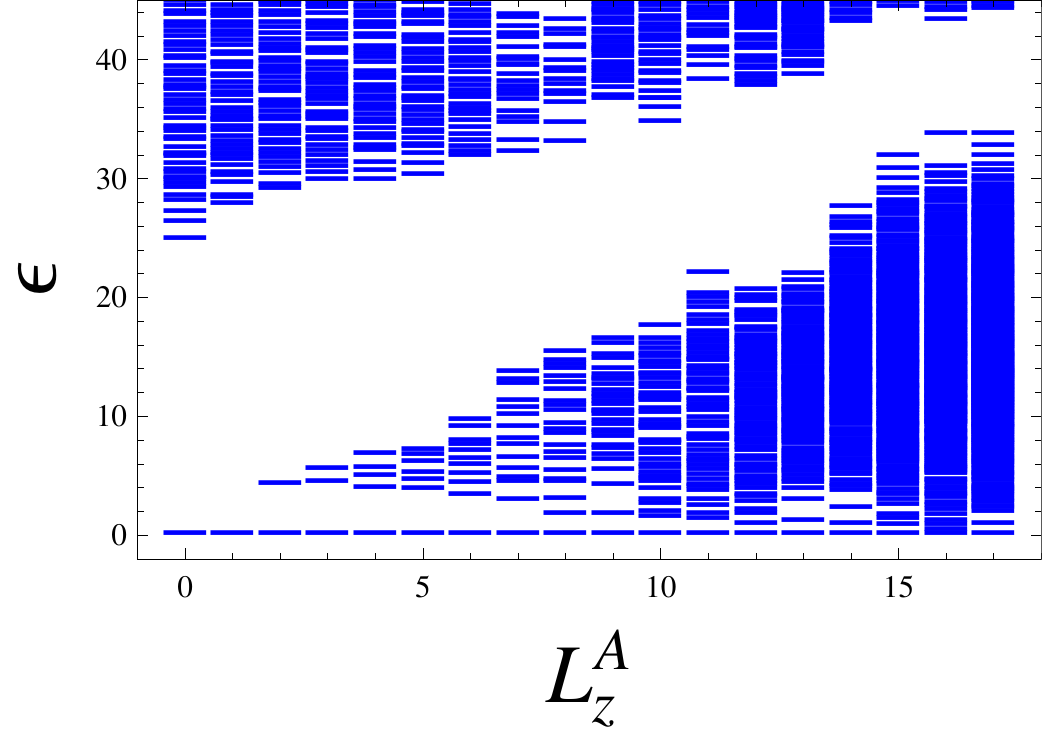}
\caption{Plot of the spectra of the matrices $M^{\dagger}M$ used in the calculation of the RSES ranks for a $\nu=\frac{1}{2}$ Laughlin state at $N=70$ and $N_A=35$ on a sphere. The negative logarithms of the eigenvalues for each $L_{z}^{A}$ block of $\rho_A$ are plotted (the largest eigenvalue is normalized to $1$ at each $L_{z}^{A}$). A clear gap is visible for each value of $L_{z}^{A}$ and the counting of large eigenvalues (at low $\epsilon$) matches the partition numbers.}
\label{fig:Laughlin13}
\end{center}
\end{figure}
Fig.~\ref{fig:Laughlin13} shows the spectra of the matrices $M^{\dagger}M$ 
for the Laughlin state at $\nu=1/2$ at different values of $L_{z}^{A}$, in the way in which entanglement spectra are usually presented ($L^A_z$ is given relative to the state with the lowest $L^A_z$, analogously to $\Delta L_z$ above). A clear gap is visible for each value of $L_{z}^{A}$ and the counting of large eigenvalues matches the partition numbers. Of course this plot is not a plot of the true RSES. Repeats of the RSES rank calculation with different initialization of the random number generator will yield different, but similar looking plots. 

For the bosonic and fermionic MR states, we also computed the RSES at $N=70$ and $L^A_z\le 20$, obtaining the expected CFT countings~\cite{Milovanovic96}. We also checked that, by changing $N_A$ or by including localized quasiholes, one may obtain the CFT countings for different topological sectors, in analogy to the results for the OES\cite{Li08,Papic11}. 

We have computed the PES ranks beyond the universal CFT numbers, for $\nu=1$, $\nu=\frac{1}{2}$, and $\nu=\frac{1}{3}$ and for the fermionic and bosonic MR states, for many values of $L^A_z$ at $N=50$ and for all $L^A_z$ in smaller systems, 
obtaining the expected finite size countings~\cite{Read96,Hermanns11}. 

At integer filling $\nu\in\mathbb{N}$ the OES ranks are trivial but the RSES ranks are the full ranks of the angular momentum sectors of the Hilbert spaces for $N_{A}$ particles in $\nu$ Landau levels.
For $\nu=1$, the ranks are just the partition numbers (\ref{count_Laugh}) in the thermodynamic limit, in agreement with Ref.~\cite{Rodriguez09}, which treated the $\nu=1$ system analytically in a second quantized formulation. For general integer filling $\nu=p$, the ranks depend on the residue of $N_A$ modulo $p$. E.g.~for $\nu=2$, a generating function for the thermodynamic countings is
\begin{equation}
\label{eq:nu2count}
Z_{\nu=2} = \left(\sum_{m\in\ZZ} q^{m^2-s m}\right)\left(\prod_{k > 0}  \frac{1}{1 - q^k}\right)^2, 
\end{equation}
where $s=N_{A} (\mathrm{mod}~2)$.
Similar formulas may be obtained for higher integer $\nu$. In fact, it can be shown that 
for $N_A=0~({\rm mod}~\nu)$ the ranks always equal the numbers of $\nu$-colored generalized Frobenius partitions~\cite{Andrews84}. 
We checked that our method reproduces these numbers for $\nu\le 3$, for systems of around $30$ particles and necessarily for modest angular momenta, as the ranks grow quickly.
%

Finally, we apply our method to the Jain state at $\nu=\frac{2}{5}$.
Jain states~\cite{Jain89} have so far produced serious challenges in ES calculations because they lacked a clear `entanglement gap' at the accessible sizes. While the numerical method described here calculates RSES ranks, we can expect the low lying parts of the spectra of the matrices $M^{\dagger}M$ to converge to yield the low lying parts of the true RSES, as the size of the matrices $M$ is increased.  Fig.~\ref{fig:Jain25} shows the spectra of the matrices $M^{\dagger}M$ at low $L_{z}^{A}$ for the $\nu=\frac{2}{5}$ CF state at $N=20$ and $N=22$, obtained with matrices $M$ which are large enough to allow the pattern of entanglement energies to emerge. Note that so far, no ES for this state has been published for $N>10$, due to the difficulty of obtaining the wave function in angular momentum space at large $N$. At low $L_z^A$ values, we observe several branches of edge states in the RSES. Each branch has ranks $1,2,5,10,\ldots$, consistent with the counting for a pair of non-interacting Luttinger liquid edges. The total rank at each $L_z^A$ equals the dimension of the spaces of CF states with $N_A$ CFs in $2$ CF Landau levels at the given value of $L_{z}^{A}$. The ranks are in general smaller than those for $\nu=2$ because some CF states disappear in the lowest LL projection. Note that the CF LLs we used for the $A$ system are the same as those for the full system. This is natural for the low $L_{z}^{A}$ edge states of the RSES. For a description of the full PES, one should consider CFs 
at a higher effective flux, 
see Ref.~\onlinecite{Sterdyniak11_PRL} for details.
We further checked that the space of eigenvectors of $M^{\dagger}M$ at each $L_z^A$ has very close to unit overlap with the space spanned by the vectors $v_i=\phi^{CF}_{i}(Z^{j}_{A})$, where the $\phi^{CF}_{i}$ are the CF trial wave functions for system $A$ and the $Z^{j}_{A}$ are the configurations of the particles in subsystem $A$ that were used in the construction of $M$. 
It appears that the low lying RSES of this CF state is excellently described by the CF trial wave functions for the edge excitations of the $A$ system.    
In future work, we intend to expand our results on CF states and on the calculation of entanglement energies.
%
 


\noindent \textbf{Acknowledgments:} The authors thank Nicolas Regnault for illuminating discussions. SHS and JKS acknowledge the support and hospitality of the Aspen Center for Physics. JKS and IDR were supported by SFI Principal Investigator award 08/IN.1/I1961. SHS is supported by EPSRC grant EP/I032487/1.

\noindent \textbf{Note added:} As this work was being completed, we became aware of research studying the RSES by A.~Sterdyniak, A.~Chandran, N.~Regnault, B.~A.~Bernevig and P.~Bonderson~\cite{Sterdyniak11_arxiv},  and by  J.~Dubail, N.~Read and E.~H.~Rezayi~\cite{Dubail11}.


\end{document}